\documentclass[twocolumn,prl,superscriptaddress,floatfix,aps,showpacs]{revtex4}

\usepackage{bm}
\usepackage{psfig}
\usepackage{epsfig}
\usepackage{graphicx}
\usepackage{amsfonts,amssymb,amsmath}
\begin{document}

\title{Superfluidity of Dense $^4$He in Vycor}
\author{Saad A. Khairallah}
\affiliation{Dept. of Physics, University of Illinois at
Urbana-Champaign, Urbana, IL 61801, USA}
\author{D. M. Ceperley }
\affiliation{Dept. of Physics, University of Illinois at
Urbana-Champaign, Urbana, IL 61801, USA} \affiliation{NCSA}

\begin{abstract}

We calculate properties of a model of $^4$He in Vycor using the
Path Integral Monte Carlo method. We find that $^4$He forms a
distinct layered structure with a highly localized first layer, a
disordered second layer with some atoms delocalized and able to
give rise to the observed superfluid response, and higher layers
nearly perfect crystals. The addition of a single $^3$He atom was
enough to bring down the total superfluidity by blocking the
exchange in the second layer. Our results are consistent with the
persistent liquid layer model to explain the observations.  Such a
model may be relevant to the experiments on bulk solid $^4$He, if
there is a fine network of grain boundaries in those systems.

\end{abstract}

 \maketitle

There is a long history of experiments of helium absorbed in
porous media carried out to probe the response of superfluidity to
disorder \cite{reppy}. Recently, Kim and Chan reported the
observation of a supersolid Helium phase, in porous Vycor
\cite{chan1} and in bulk solid $^4$He \cite{chan2}. A
supersolid \cite{al-leg-chest} is a proposed phase of a quantum
system in which long-range crystalline order and superfluidity
coexist.  We focus here on the measurements of the Helium-Vycor
system.

The occurrence of supersolid behavior in Vycor, a disordered
porous glass, could be understood from the properties of Vycor.
One explanation is that the complex Vycor geometry stabilizes
mobile defects which then undergoe BEC at low temperatures
$\approx$ 0.2K.  However, since the phenomenon is observed in bulk
solid $^4$He it is not the Vycor pores that are essential.
Although, Kim and Chan pressurized their cell to 60 Bars,
substantially above the estimated freezing pressure for helium in
Vycor of 40 bars, even if most of the helium is solid, it is not
clear if there remains a liquid film  (the persistent liquid layer
or PLL) near the Vycor-helium surface. The film could arise from
the mismatch of lattice parameters as the density of $^4$He varies
from the center of the pores to their surface.

Even though there have been many experimental studies of helium in
Vycor, there have been few microscopic calculations of the
detailed microscopic structure of this system. Here we report on a
model of the helium-Vycor system and calculate its properties with
the Path Integral Monte Carlo(PIMC) method \cite{pimc}. PIMC can
calculate exact thermodynamic properties of bosonic systems such
as $^4$He at non-zero temperature by sampling the thermal density
matrix $\rho\equiv e^{-\beta H}$, with $\beta=1/k_{B}T$ and $H$
the Hamiltonian. An explicit expression for the density matrix is
obtained by expanding into a path and approximating the higher
temperature
density matrices. 
Bose statistics are obtained by symmetrizing the density matrix
$\rho_{Bose}(R,R';\beta)=\sum_{P}\rho(R,PR';\beta)/N!$. PIMC
proved accurate in studying properties in the normal liquid,
superfluid
and crystal phase \cite{pimc}. 
In contrast to methods based on trial wavefunctions, in PIMC only
the Hamiltonian enters, so no biases are made on the structure of
the many-body system.
\begin{figure}[ht]
\centering
\begin{minipage}[r]{.4\textwidth}
\includegraphics[width=\textwidth]{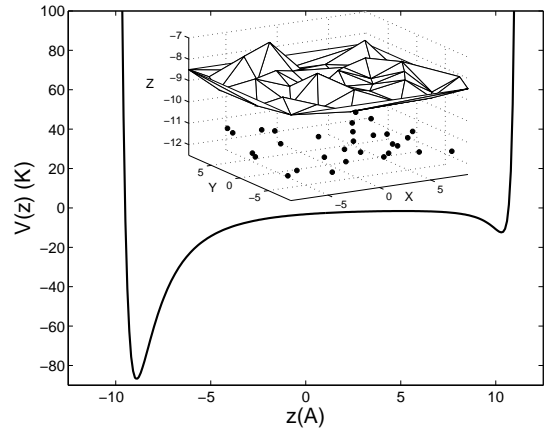}
\end{minipage}
\caption{\label{model}The external potential $V(z)$ experienced by
the helium atoms in ``cell 2''. The Vycor is on the left side
(z=-12.5\AA) and solid helium on the right side (z=12.5\AA).
The inset is a 3D representation of the rough Vycor surface: the
black dots are the positions of the Vycor impurities placed
randomly at 1\AA \space away from the Vycor wall. The rugged
surface shows the positions of the helium atoms (located at the
vertices) in the first layer.}
\end{figure}

Vycor's complex geometry is difficult to simulate directly. Under
the transmission electron microscope, Vycor shows interconnected
pores with diameters between 60\AA \space and 80\AA \space and a
length of ~300\AA, with a narrow distribution of pore sizes.
Current simulation techniques treating all the atoms  with PIMC
are unable to simulate even a single pore (containing roughly
30000 atoms). It is thought that above the freezing pressure (40
bars) a $^4$He crystal will occupy the center of a pore
\cite{brewer}. Previous PIMC calculations have found that a
perfect crystal will not have a superfluid response at long
wavelength \cite{C181}. Making this assumption, we study only the
region near the surface of a pore and model it with the slab
geometry: our simulation cell is periodic in both the x and y
directions. In the negative z-direction there is a wall,
representing bulk Vycor on top of which are placed Vycor
``particles'' in random positions 1 \AA \space above the flat
Vycor surface. These particles serve to make the surface rough and
break translational invariance. In the positive z-direction is a
wall which models bulk solid Helium at a pressure of 62 bars(see
Fig.~\ref{model}) .

The main input to PIMC is the interaction potential between
various particles. We assume that the Helium-Helium interaction is
given by the Aziz \cite{aziz} potential which has been used within PIMC to
study numerous other properties of helium and gives energies
accurate to about 1\% of the bulk binding energy. The potential
between helium and the upper wall was derived by integrating an
approximate LJ 6-12 potential ($\epsilon = 10.22K$ and
$\sigma=2.556 \AA$) over the volume $z> 11.7\AA$ (cell 1)
resulting in a LJ (3-9) potential.

We also assume helium-Vycor surface interaction is a LJ (3-9)
potential \cite{pot}: $ V(z)= \frac{D}{2} [( \frac{z_e}{z})^9-3(
\frac{z_e}{z})^{3}]$. Since Vycor glass, SiO$_2$, should behave
similarly to MgO \cite{Vycorpot}, we pick the well depth to be
$D=-86.9K$, and the range of the attraction, $z_{e}=3.6\AA $. In
order to pin the helium crystal in the xy plane and model the
roughness of the Vycor, we add Vycor ``particles'' in random
positions 1 \AA \space above the wall. The interaction between the
helium atoms and the Vycor particles is determined by demanding
that a complete layer of the particles give the same LJ
(3-9)potential. Experiments \cite{levitz} show that the roughness
in Vycor is on the scale of 0.8nm, not very different from what we
have assumed. Figure ~\ref{model} shows the wall potential and the
surface roughness.

We set the helium density to match the experimental conditions by
adjusting the number of helium atoms and the total area in the xy
plane so that the helium density in the topmost layer matches that
of solid $^4$He at a target pressure close to 62 bars. We start
the simulation with 221 atoms placed in 7 layers in an hexagonal
closed packed solid phase. Each layer contains 30 atoms except for
the first layer placed at the strongly attractive Vycor wall
potential (Fig.~\ref{model}). We have performed extensive
simulations with two geometries, denoted as Cell 1 (221 He atoms
with a box $17.75\times18.45\times23.4$\AA) and Cell 2 (221 atoms
with a box $17.25\times17.93\times25$\AA). Cell 1 is roughly
stress free, while Cell 2 provides us with a way to look at the
model under anisotropic stress.

\begin{figure}
\begin{minipage}[r]{.45\textwidth}
\includegraphics[width=\textwidth]{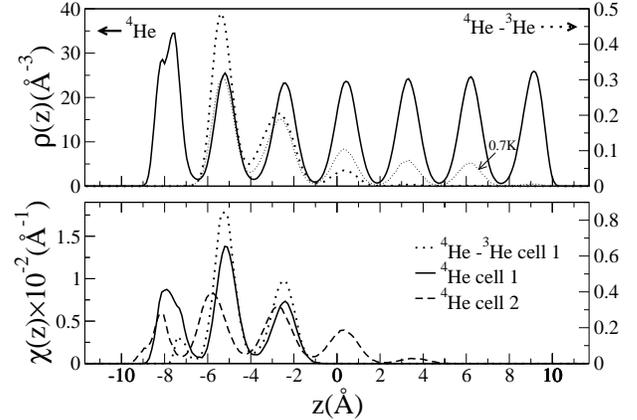}
\end{minipage}
\caption{\label{densities}The density and superfluid density as a
function of $z$. The left scale (solid line) corresponds to pure
$^{4}$He and the right to the mixture: one $^{4}$He atom replaced
by a $^{3}$He atom. Top: The density in the z-direction for cell 1
at 0.2K. The split peak in the first (leftmost) layer is due to
the rough Vycor surface.  The dotted lines show the $^{3}$He
density at 0.2K and 0.7K. Bottom: Local superfluid density of
$^4$He determined by recording which layers the winding paths
visit; Eq. (3).}
\end{figure}

The density in the z-direction (see Fig.~\ref{densities}) shows a
distinct layered structure of $^{4}$He. The density increases as
we approach the Vycor wall because of the stronger attraction of
the potential well. The Vycor particles distort the shape of the
density peak in the first layer because there are binding sites at
different values of $z$. We find a perfect crystal is stable in
the upper portion of the cell with a lattice constant of 3.55
\AA\, in cell 1 and 3.45 \AA\, in cell 2, at the target pressure
of 62 bars. We estimate the pressure by calculating the density
per layer from fig.~\ref{densities} and comparing it with the
experimental equation of state \cite{state}.

\begin{figure}[ht]
\centering
\begin{minipage}[l]{.45\textwidth}
\centering
\includegraphics[width=\textwidth]{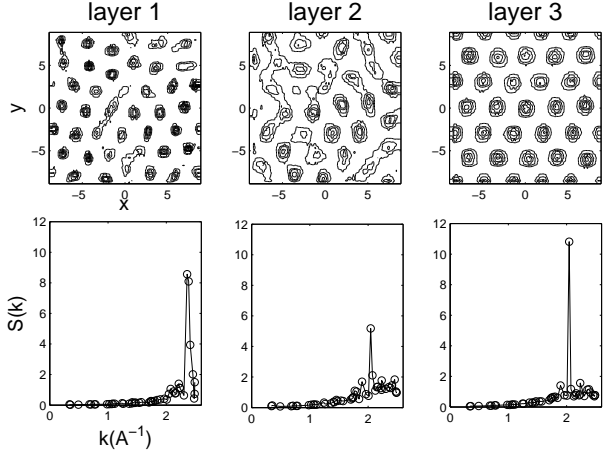}
\end{minipage}
\caption{\label{sofkcontour}Top: Contour plots of the helium
density in the x-y plane in the first three layers at 0.2K for
cell 1. In the first layer above the Vycor, the atoms are pinned
by the strong Vycor interaction, in the second layer they are
relatively delocalized, whereas the higher levels show a density
distribution characteristic of bulk solid $^4$He. Bottom: The
layer-specific structure factor. The first layer has an amorphous
structure, the second is still distorted but with a much smaller
peak, while  the third and higher layers have a single large peak
at k$_{o}$=2.04\AA$^{-1}$ characteristic of a 2D quantum solid.}
\end{figure}

We determine the spatial ordering within a layer with the
structure factor in the x-y direction
\begin{equation}
S_n(\bold k)=\frac{1}{N} \langle \rho_n(\bold k)\rho_n(-\bold k)
\rangle
\end{equation}
where $\rho_n(\bold k)= \sum_{i=1}^{N}\Theta(z_i\in n)\exp(i\bold
k \cdot \bold r_{i})$ is the Fourier transform of the density
within layer $n$ and $\bold k=(k_{x},k_{y},0)$. We can see the
signature of a solid from the peak of $S(\bold k)$ around
$k_{0}$=2.04\AA$^{-1}$ as shown in fig.~\ref{sofkcontour} for cell
1; the peak clearly shows an hexagonal structure for layers three
and above.  The density profiles within a layer at the top of the
figure confirm this interpretation. However, not all of the layers
are solid. In fact, layer 1 is solid-like with the helium atoms
well-localized but with their mean positions determined by the
underlying disorder. Layer 2 is more disordered, and the atoms are
out of register with the first layer. Because the second layer
density is lower, the atoms are much less localized and, as we
shall see, are able to become superfluid. Layers three and above
are quantum solids, mostly free of defects. This is relatively
independent of pressure, because of the strong Vycor potential
shown in Fig. (1). We note that recent neutron scattering
experiments \cite{wallacher} support the co-existence of solid-like
and liquid-like layers at these pressures.

\begin{figure}[ht]
\centering
\begin{minipage}[l]{.45\textwidth}
\centering
\includegraphics[width=\textwidth]{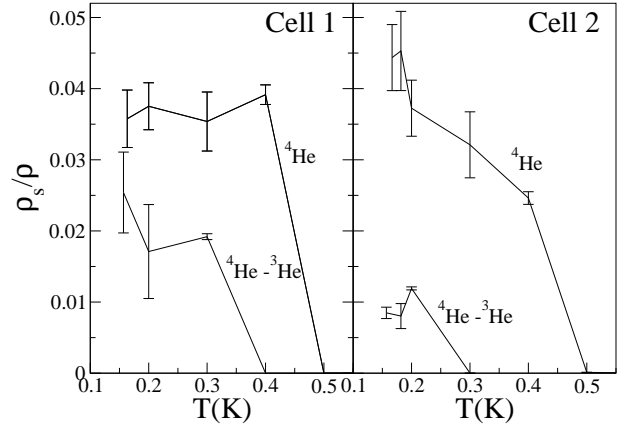}
\end{minipage}
\caption{\label{SF}The superfluid fraction $\rho_s/\rho$ vs.
temperature. The upper curve is for pure $^4$He. The lower curve
is for the system with a $^4$He atom replaced by a $^3$He atom.}
\end{figure}

The superfluid fraction is computed in PIMC from
the mean squared winding number \cite{pimc4},
$\rho_s/ \rho = m\langle W^2 \rangle/(2\beta \hbar^2
N)$ where $N$ is the number of Helium atoms and $\bf{W}=\sum_{i,k}
(\bf{r}_{i,k}-\bf{r}_{i,k+1})$; the sum is over particles $i$, and
time slices $k$.  The superfluid fraction increases as we lower
the temperature as shown in fig.~\ref{SF} and approaches values
of about 4\% below 0.3 K.

To find the spatial distribution of the superfluid density, we
divide the winding number estimator into local
contributions that sum to the total superfluid density \cite{eric}. The
superfluid density as a function of the distance above the Vycor
wall $\chi(z)$ is:
\begin{equation}
\space \chi(z)=\frac{\sum_{k_{slice},c_{cycle}}W^{2}_{k,c}
\delta(z_{k,c}-z)}{4\lambda \beta N}.
\end{equation}

In fig.~\ref{densities}, we show $\chi(z)$ for T=0.2K in cell 1.
One can see the layered structure of the density. Layer 2 has the
largest superfluid component. Layer one contributes because some
atoms
sit close to layer 2. 
Layer 3 is also active. The superfluid response goes to zero above
layer 3. However, this decay is slower in cell 2 where the two
additional layers (4 and 5) still contribute to $\rho_s/ \rho$.

To compare with experiment we must make two corrections; first
that our model has a larger fraction of atoms near the Vycor
surface than experiment, and second that our cell has no
tortuosity: the experimental path length for a superflow is
greater than the straight line distance. We correct for the
difference in Vycor surface area by assuming that the
superfluidity is confined to a finite distance above the Vycor so
that the effective number of atoms that do not respond to the
moving boundaries is: $N_s = \sigma A$ where $A$ is the Vycor
surface area (actually the surface area of the mobile layer). We
find that the maximum number of superfluid atoms per unit area is:
$\sigma \approx 2.7 nm^{-2}$. Then for a Vycor sample with
experimental surface area per unit volume of $a=0.2 nm^{-1}$ the
relative superfluid response would be $ a \sigma_s/\mu \rho
\kappa$ where $\mu=0.3$ is the experimental Vycor sample pore
fraction, $\kappa=5$ is the tortuosity, and $\rho=31.66 nm^{-3}$
is the solid number density. Using the experimental parameters of
Vycor, we would predict a measured superfluid fraction of $0.011$.
In fact, Kim and Chan measure a value two times smaller than this.
One important effect missing in our calculation is that the
supercurrents must tunnel through various weak links, an aspect
not present in our calculations because of the limited extent of
the cell in the x-y directions. In addition, we have not taken the
thermodynamic limit, though it is not obvious how to do this
without a better model of the Vycor. We also note that the 
model is lacking important aspects of
disorder, for example, that induced by the curvature of the pore
and possible grain boundaries between the pores. Given these
uncertainties, Kim and Chan's measured values are well within the
range expected from our calculations.

Recent experiments \cite{day} do not find any indication that the
liquid layer can flow in response to a pressure gradient. It is
possible that there is an ``insulating'' ({\it e.g.} defect-free,
see below)layer  of solid $^4$He which forms at the surface of the
Vycor samples and blocks the superflow.

Kim and Chan also studied the effect of $^3$He impurities on the
measurement of the superfluid density. Accordingly, we replaced  a
single $^4$He with a $^3$He atom. (Note that we do not need to
consider fermi statistics for a single fermion). This corresponds
to roughly 0.3\% concentration of $^3$He (assuming our cell size
is $\approx 2/3$ of the pore). Experimentally, this concentration
was  enough to destroy any ``supersolid'' response. With local
moves the system was slow to reach equilibrium: we used a move
that swapped the identity of the impurity with a random $^{4}$He
atom. PIMC with the new procedure quickly reached equilibrium.

The impurity, being lighter than $^4$He, has a larger kinetic energy 
and hence a larger atomic volume. This extra
space is available in the second liquid-like layer. Also, the $^3$He atom does
not exchange with the $^4$He and thus reduces the superfluid
density in its neighborhood. The density profiles show that the
$^3$He impurity preferentially goes to the same site where the
superfluidity is maximized (Fig. \ref{densities}). At higher
temperatures, it tunnels to other layers with an excitation energy
of 0.8K. As the temperature is lowered, two competing effects take
place. The $^{4}$He atoms nearest to the Vycor surface exchange
resulting in superfluidity. However, the $^{3}$He atom migrates
towards the most superfluid layers, hence reducing the superfluid
response and a shift towards a lower transition temperature, see
fig. \ref{SF}.  It is plausible that in a larger cell, an even
smaller concentration of $^3$He could pinch off the winding
exchanges (or supercurrent) by going to choke positions not
present in our small cell, thus giving better agreement with the
experimental findings of a critical $^3$He concentration of 0.1\%.

Kim and Chan \cite{chan1} mention two pieces of evidence to argue
against the liquid film interpretation.  First, the observed
temperature dependance of the superfluid density is unlike that of
films. However, the films under pressure are totally enclosed
within a solid and they are not like films forming an interface
between a solid and a vacuum; as we have seen, there are low
energy excitations giving rise to pronounced temperature effects,
not present in the later case. Also, the connectivity whether
primarily 2D or 3D, of these solid-enclosed films could be
different; it is likely that they are gossamery, as opposed to the
robust films resulting when the pores are only partially filled.
The second effect mentioned by Kim and Chan is the poisoning by
small amounts of $^3$He. This does not happen in free films
because the $^3$He atom is above the plane of the $^4$He film in a
delocalized state and thus is not effective in preventing $^4$He
exchanges.

The liquid-layer scenario can be used to give insight into
the other porous media and bulk helium experiments. Let us
assume that the observed superfluidity is due to surface
superfluidity as proposed by Burovski at al. \cite{burovski} and
that the helium microcrystals are roughly spherical with a diameter of
$R$, giving a surface to volume ratio of $a=3/R$. Arguing as
before, to obtain the observed 2\% superfluid response of bulk
$^4$He, we must have that $R=3 \sigma/(\kappa\rho_s)$ where
$\sigma$ is the superfluid density per unit area at the crystal
interfaces. It is reasonable to expect that $\sigma$ is
considerably larger than at the interface between helium and
Vycor, since the pressure exerted by the silica increases the
density and hence decreases the mobility of the helium atoms. Let
us assume that there are 2 complete superfluid layers at the
interface giving $\sigma \approx 20 nm^{-2}$. Given that the grain
boundaries are larger and less fractal, we might expect that
$\kappa \approx 2$. Using the experimental $\rho_s$ for bulk
$^4$He, we find the average grain size, $R=50 nm$.  It is quite
likely that $^3$He will stabilize the grain boundaries, thus
explaining how such a small concentration of $^3$He can affect the
response. Such speculation need to be confirmed by performing
experiments on much better crystals and measuring the density and
the sizes of grains.

Our results show superfluidity is localized in specific layers of
$^4$He  above a Vycor surface. We obtain a superfluid response
about 2 times what is observed, but the difference is likely due
to the very small simulation cells we used that do not have the
full range of the random disorder and crystal defects. We also
found that $^{3}$He impurities gravitate to the same spatial
locations as the superfluid density, thus poisoning the effect.
Based on these simulations, the persistent liquid layer
interpretation of the Kim-Chan experiment seems not to be ruled
out. Further studies with larger cells and more realistic disorder
are needed to firm up these conclusions and to see if this
mechanism, when involving grain boundaries, could give rise to 
the phenomena in bulk solid helium.

Thanks for assistance from J. Kim, M. Chan and M. Cole. This work
was supported by NSF and the fundamental physics program at NASA
(NAG-8-1760). Computer time has been provided by NCSA and the F.
Seitz Materials Research Lab. (US DOE DEFG02-91ER45439 and NSF
DMR-03 25939 ITR), at the U. of Illinois Urbana-Champaign.

\end{document}